\documentclass[conference]{IEEEtran}
\IEEEoverridecommandlockouts
\usepackage{cite}
\usepackage{amsmath,amssymb,amsfonts}
\usepackage{algorithmic}
\usepackage{graphicx}
\usepackage{textcomp}
\usepackage{xcolor}
\def\BibTeX{{\rm B\kern-.05em{\sc i\kern-.025em b}\kern-.08em
    T\kern-.1667em\lower.7ex\hbox{E}\kern-.125emX}}
\begin{document}

\title{V2X Enabled Emergency Vehicle Alert  System\\
}
\author{\IEEEauthorblockN{1\textsuperscript{st} Nitish Kumar}
\IEEEauthorblockA{\textit{Dept. of Electrical Engineering} \\
\textit{Indian Institute Technology,}\\
Hyderabad, India \\
ee22mtech02005@iith.ac.in}
\and
\IEEEauthorblockN{2\textsuperscript{nd} Hershita Shukla}
\IEEEauthorblockA{\textit{Dept. of Integrated Sensor System} \\
\textit{Indian Institute Technology,}\\
Hyderabad, India \\
is21mtech14005@iith.ac.in}
\and
\IEEEauthorblockN{3\textsuperscript{rd}  Prof. P. Rajalakhsmi}
\IEEEauthorblockA{\textit{Dept. of Electrical Engineering} \\
\textit{Indian Institute Technology,}\\
Hyderabad, India \\
raji@ee.iith.ac.in}
}
\maketitle

\begin{abstract}
Today's major concern in traffic management systems includes time-efficient emergency transports. The awareness of environment and vehicle information is necessary for the emergency vehicles as well as the surrounding commercial vehicles that might be driven by inexperienced drivers to act accordingly if they both interact. The information exchange should be quick and accurate along with how much interactive the alerting system is with the drivers. Therefore, technologies like V2X-based alert systems can deal with such emergency situations and hence prevent potential health or social hazards. An alerting system as a part of a smart-connected city is proposed in this paper. The Dedicated Short Range Communication (DSRC) based system has tried to cover the major domain of information about misbehaving vehicles, any pedestrians on the road, and information about the emergency vehicle itself. The commercial vehicle also will have a similar alert system as an application of V2V and V2I. Further in this paper, a realtime monitoring system was developed using grafana dashboard which will be installed in the area's base station to monitor the vehicles in that area. 
\end{abstract}


\section{Introduction}
 Emergency vehicles include ambulances that are responsible for the transportation of any patient that needs an immediate response during life-threatening situations. Vehicles like fire brigades and police cars also fall into the same category.Due to drivers' ignorance of what to do when they encounter an ambulance on the road, one in ten patients in India pass away on the route to the hospital, according to the Radhee Disaster and Education Foundation \cite{b1}. In Mumbai, 1500–2000 ambulances are typically off the road each day, despite the fact that this worsens the patient's health and frequently results in death. A delay in receiving medical care results in almost 24,012 deaths daily, according to the National Crime Records Bureau.\\
 \\
 According to a study of AIIMS \cite{b2}, the major leading causes of Death and disability-Adjusted Life Years (DALY) globally were the conditions with potential emergent manifestations which primarily occur due to untimely response of emergency vehicle support The detailed analysis is done in figure 1.\\
 \begin{figure}[htbp]
\centerline{\includegraphics[width=0.5\textwidth]{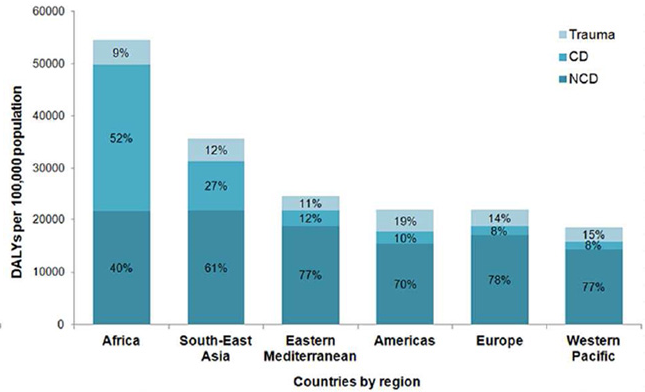}}
\caption{DALYs per 100,000 population attributable to emergency conditions, Distribution of deaths was similar. Abbreviations: non-communicable diseases (NCDs), communicable-diseases (CDs), disability-adjusted life years (DALYs).\cite{b3}}
\label{fig}
\end{figure}
\\
 This project has been made to support the smooth movement of these vehicles hence, we have developed an alert system with DSRC based communicative hardwares (OBUs and RSUs) that can exchange information directly amongst the vehicles (V2V), vehicles to smart poles (V2I) and also from smart poles to the base station for overall monitoring of the region.
\section{Literature Review}
In the case of emergency vehicle alert system proper alert at proper time is very important in real-time. Due to traffic congestion, emergency vehicles are unable to get to the scene in time to provide the patient or victim with the necessary care, which results in fatalities. The researchers are working on this problem statement and came up with multiple solutions. This dissertation suggests a framework for automatically creating "Emergency Corridors" to allow for the quick clearing of emergency vehicles in order to deal with such emergency scenarios \cite{b4}. Using V2I communication, the traffic signal controller dynamically suspends the regular passage of traffic to give the emergency vehicle's approach the green light, while V2V communication directs the vehicles in front to move aside. They have implemented their work as a simulation result in VEINS framework with combination of SUMO and OMNeT++. Many academics have suggested an emergency vehicle clearance system based on acoustics for quick clearance of the emergency vehicle \cite{b5}\cite{b6}. The emergency vehicle employs a sound-based alerting system to notify the traffic intersection and other individuals. In this kind of system, it is exceedingly challenging to determine the precise location and direction of travel of emergency vehicles at the traffic intersection. Eltayeb \cite{b7} suggests an internet of things (IoT) based traffic management system for emergency vehicle clearance. This technology uses GPS to forecast traffic congestion and location. For image-based traffic monitoring \cite{b8}\cite{b9}\cite{b10}, several strategies have been put forth. This system uses camera modules that are positioned at traffic lights to record frame-by-frame images of moving vehicles. Deep Neural Networks are then used to identify the type of vehicle and the number of moving vehicles. Traffic congestion is assessed based on the number of vehicles. Prior knowledge of the traffic situation aids in congestion control. However, this system's performance degrades the surroundings change.\\
\\
For object and other pedestrian tracking, the tracking accuracy, tracking speed, and computational complexity of various object tracking techniques (both conventional and Deep learning-based) have been compared in this work \cite{b11}. The evaluation is based on the precision of object detection and tracking at the start, finish, or midpoint of any occlusion scenario in the drone's video. For mobile-based alerting system \cite{b12}, some Apps have been developed which is used to generate alerts about the accidents and ensures no delay in the arrival of ambulance due to road traffic. They track the location of the ambulance from Google APK and gives information to traffic police. Then traffic police needs to clear the traffic accordingly. In \cite{b13}, The author has put forth a useful technique for the early detection and alerting of potentially harmful driving behaviors. The only thing that needs to be done to implement the entire solution is to install an accelerometer and a direction sensor on a mobile phone, which will calculate the acceleration based on sensor readings and compare it to the patterns that were extracted from a real driving test to represent typical drunken driving. Results from tests using various driving styles demonstrate that the system achieves great precision and high energy economy. In \cite{b14}, utilizes the MATLAB and Simulink tools, the lane keeping assist system and ADAS characteristics were taken into account. ADAS development and qualification tests for vulnerable road users (VRU), including driving scenarios with cars and pedestrians, are examined. The effects of subject/global and ego vehicle settings are investigated using a variety of parameters, including acceleration, deceleration, perceptual reaction time, gap acceptability, and stop/go decision-making to prevent crashes. 
\section{Alert System Protocol}
\subsection{DSRC as Vehicular Communication }
One of the important technologies that will help the Intelligent Transportation System (ITS) systems to create applications based on V2V and V2I communications to increase transportation efficiency and safety is the Dedicated Shorted Range Communications (DSRC)\cite{b15}. The US FCC assigned the 5.850–5.925 GHz frequency block for DSRC. High bandwidth and rapid connection mechanisms, which are essential for automotive applications, are supported by the DSRC standard. Figure 1 shows the 75 MHz at 5.9 GHz band, 7 channel, DSRC spectrum. This spectrum was allotted for intensively used be Wireless Access in Vehicular Environments (WAVE).
\subsection{Wireless Access in Vehicular Environment (WAVE) Stack}
IEEE 802.11p is an updated version of the most widely used IEEE 802.11 Wi-Fi standard. The regular IEEE 802.11 standard cannot be used for safety-related applications in the unlicensed 2.4 GHz and 5 GHz bands \cite{b16}. The WAVE standard is intended to support automobile communications, which must meet stringent latency and capacity requirements.\\
\\
\begin{figure}[htbp]
\centerline{\includegraphics[width=0.5\textwidth]{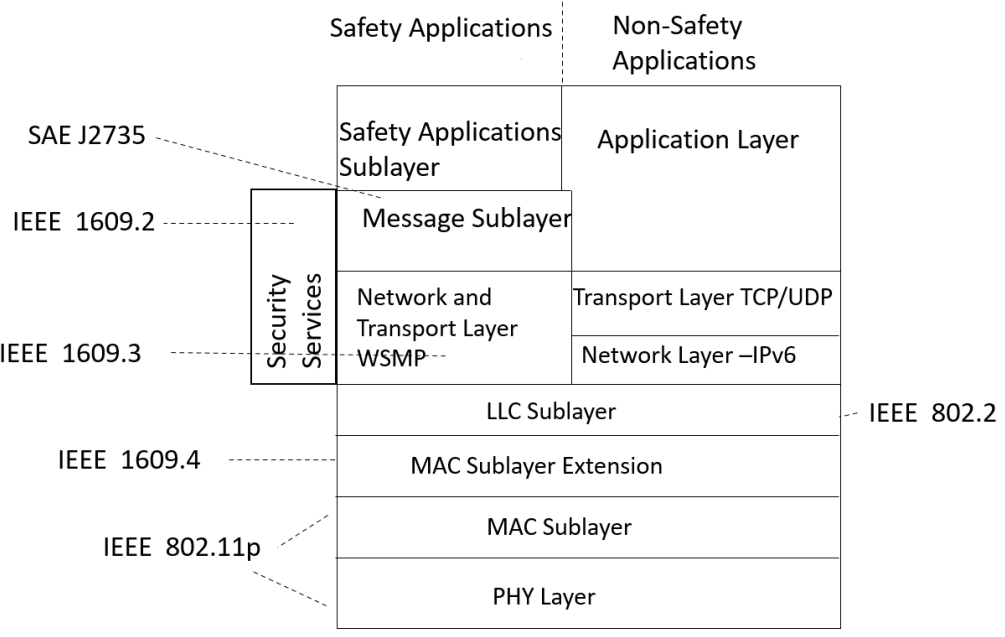}}
\caption{DSRC communication stack.}
\label{fig}
\end{figure}
Figure 2 shows DSRC WAVE communication stack. The IEEE 802.11p WAVE standard is used by the PHY and MAC levels of the DSRC stack. DSRC supports the usage of well-known Internet protocols for the Network and Transport layers, such as User Datagram Protocol (UDP), Transmission Control Protocol (TCP), and Internet Protocol version 6 (IPv6). The SAE J2735 standard is used by the DSRC stack's top layer to transmit messages about vehicle safety. The Basic Safety Message (BSM), which communicates DSRC-based information like, location, speed, and direction of the vehicle, is one of the message formats outlined by this standard. This helps in tracking the location and motion of the neighbouring cars and taking preventative measures.
\section{Alert System Hardware}
There are two major pieces of hardware for DSRC, one as the communication board on the Vehicle and the other on the infrastructure like poles, buildings, etc.

\subsection{On Board Unit (OBU)}
OBU is a component that can send and receive DSRC communications and is mounted in a car. The DSRC module’s brain is a quad-core NXP semiconductor (formerly Freescale) i.MX6 processor with 1 GB of RAM and 4 GB of NAND memory. The OBU features two IEEE 802.11p radios for DSRC communication \cite{b17}, one of which is dedicated to safety application and  other radio switches between control and other service channels to deliver DSRC based applications. It has been tested to deliver full packets without loss for upto 600m line of sight and drops to 350m without line of sight.\\
\begin{figure}[htbp]
\centerline{\includegraphics[width=0.45\textwidth]{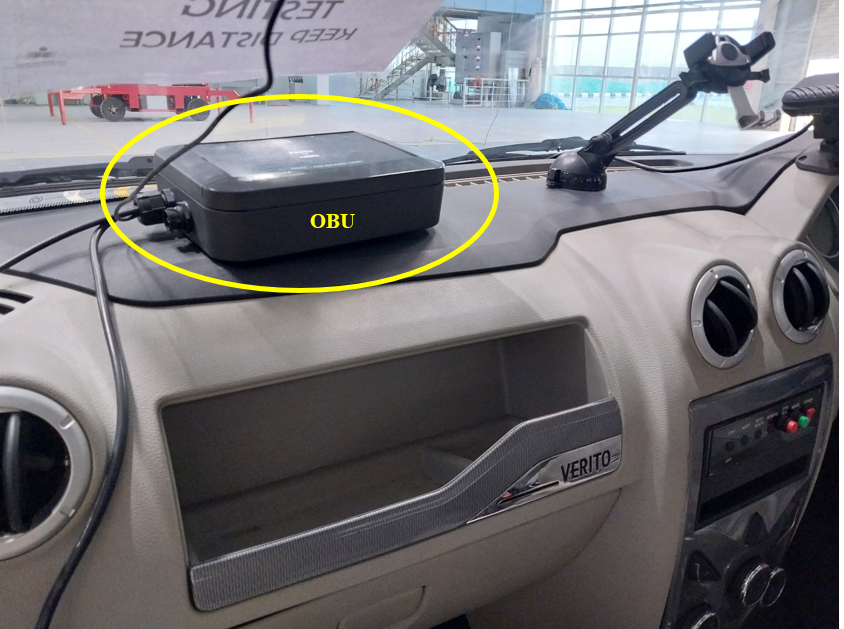}}
\caption{OBU mounted on vehicle.}
\label{fig}
\end{figure}
\\
The OBU provides Bluetooth communication which utilizes a Low Energy Bluetooth module. This connectivity can be used to setup the OBU or to communicate with adjacent devices like mobile phones. The car's CAN bus can be accessed via a CAN interface, which can then be used to extract the vehicle diagnostic messages. A micro USB OTG connector is also available on the OBU for connecting to external devices.
\subsection{Road Side Unit (RSU)}
RSUs are devices that are mounted on traffic signals or lamp-posts near roadside intersections. It gathers information of the roadside surrounding, weather, upcoming events, and other services and relays that information to the OBU and base station. Just like OBU, RSU also have an i.MX6 Quad Core processor, 1 GB RAM, and 4GB Flash \cite{b17}.
\begin{figure}[htbp]
\centerline{\includegraphics[width=0.43\textwidth]{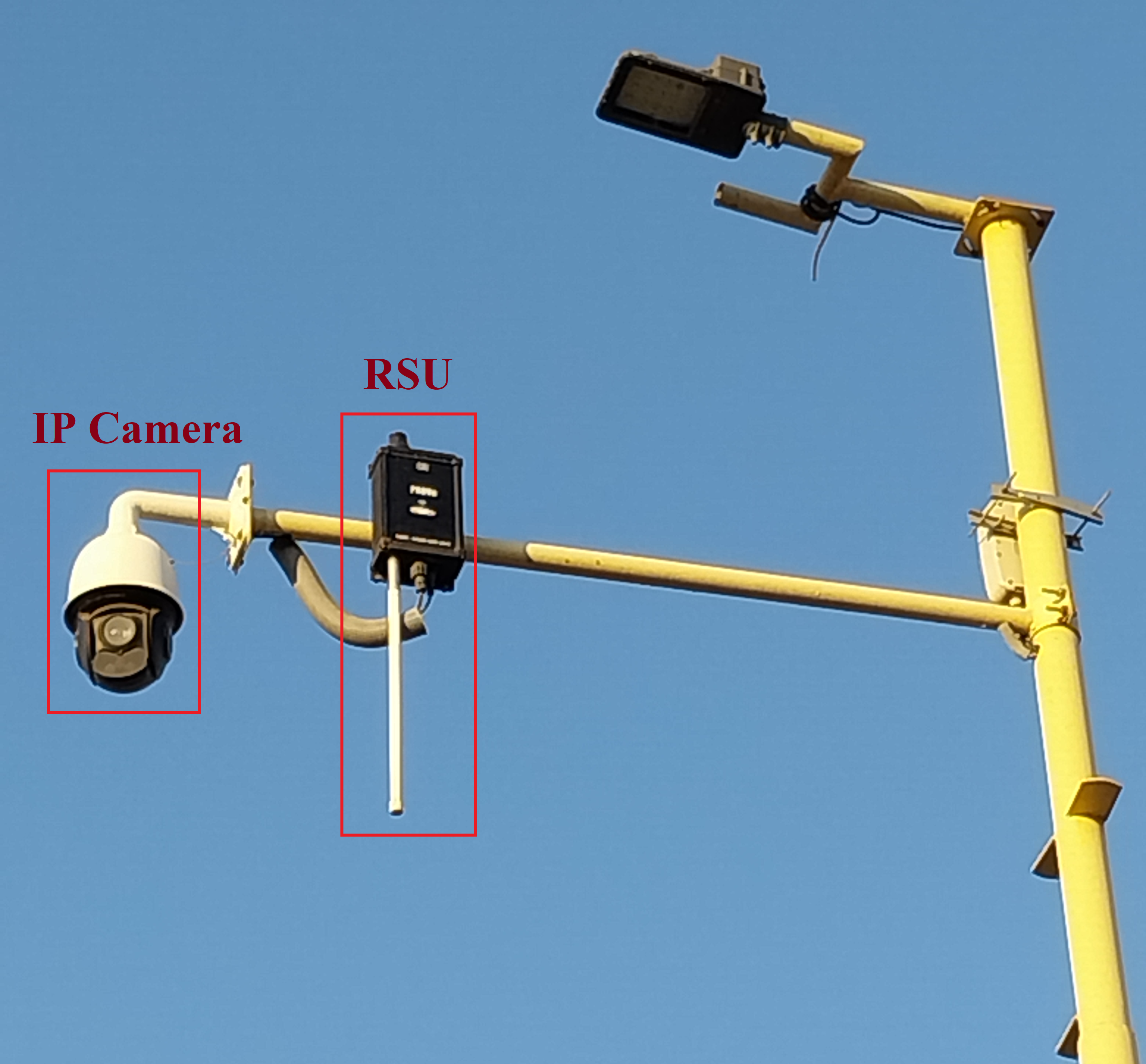}}
\caption{RSU mounted on the Traffic pole.}
\label{fig}
\end{figure}
Additionally, it functions with two DSRC radios, one for service and the other for safety. RSU also has an integrated 4G modem, Wi-Fi, and Ethernet for backbone connectivity.

\section{System Design and Implementation}
We have an embedded Linux OS integrated with DSRC Stack which is used by the OBU and RSU modules to run DSRC-based applications.
\subsection{OBU to RSU Communication}
The above describe wave stack is implemented in both OBU and RSU. When antennas of OBU comes in vicinity of RSU (or infrastructure), they start exchanging packets. Contents of the packet contain sensor data such as GPS positions, IMU data, speed of the vehicle, etc. In a similar way when a vehicle containing OBU comes in proximity to another vehicle containing other OBU then both OBU can also start exchanging packets with each other. The detailed packet study has been done and its been noticed that there is a minimum loss of packet in our OBUs and RSUs as compared to Arada's OBUs and RSUs \cite{b16}. 
\\
\begin{figure}[htbp]
\centerline{\includegraphics[width=0.5\textwidth]{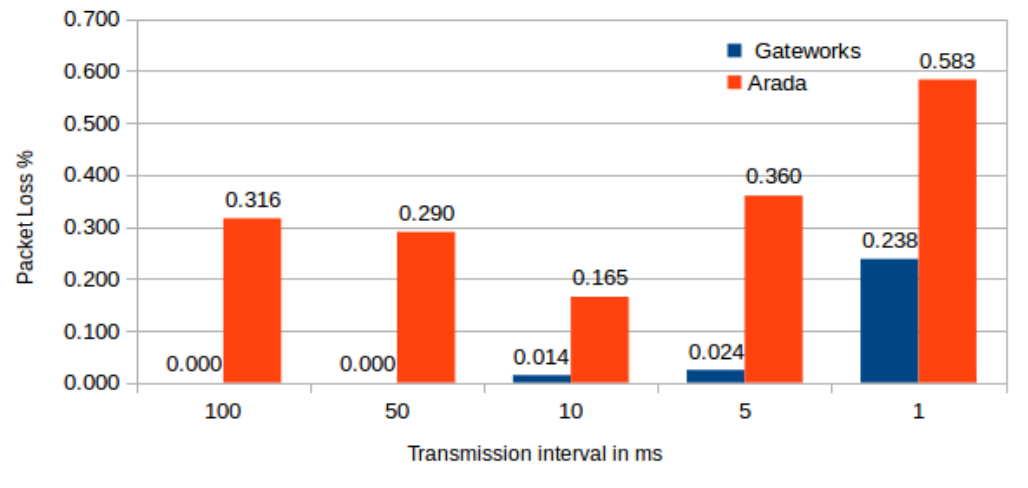}}
\caption{Percentage loss of packets in setup1}
\label{fig}
\end{figure}
\\
\begin{figure}[htbp]
\centerline{\includegraphics[width=0.5\textwidth]{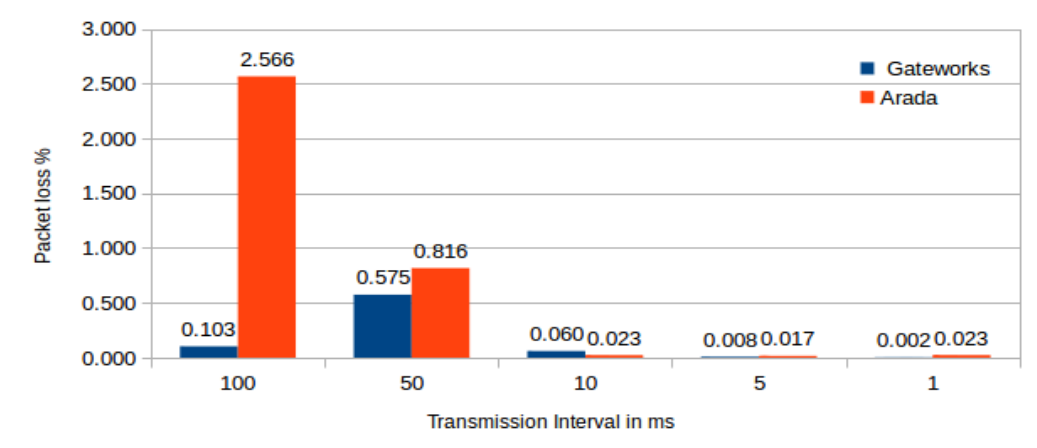}}
\caption{Percentage loss of packets in setup2}
\label{fig}
\end{figure}
\\
The Gateworks development board was used to implement the created stack. Later, we tested how well the built stack performed when used with the ARADA stack (a commercially available OBU and RSU platform) for the creation of V2V and V2I applications) \cite{b18}.\\
\\
Every OBU and RSU contains a specific ID. During transmission of packets, RSU/OBU ID is also attached in the string which is used to differentiate between emergency vehicles and normal commercial vehicles. Figure 7 describes the packets received at one particular OBU from different RSUs and OBUs with all sensor data present in that hardware.
\begin{figure}[htbp]
\centerline{\includegraphics[width=0.5\textwidth]{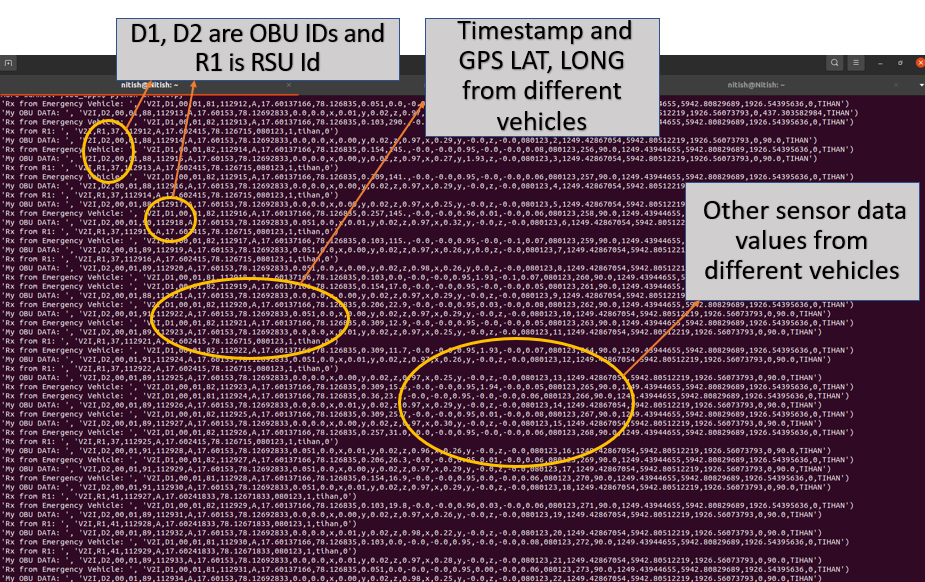}}
\caption{Different Sensor data with OBU and RSU ID receiving at a vehicle's OBU}
\label{fig}
\end{figure}
\subsection{Android App Development}
For user-friendly interface, we have developed an App that can be installed on HMI of vehicles and is connected with Bluetooth of OBU mounted on that vehicle. The OBU circuitry has an RN-4020 low Energy Bluetooth module which enables it to interact with the HMI, each OBU associated with each vehicle has its own character/service ID Universally Unique Identifiers (UUIDs), these service IDs are sent in the form of alerts to the app. This app contains all the alerts necessary for a vehicle that is to be displayed on HMI. So, whenever an OBU gets any alert from the surrounding environment (RSU or OBU), the same alert is displayed on the App via Bluetooth connectivity. There are two Apps showing a specific alert for emergency vehicles and other for normal commercial vehicles. Some of the alerts that are displayed on the app are over-speeding, collision aware, pedestrian detection, etc. Figure 8 shows a screenshot of our android App used in emergency vehicle.
\begin{figure}[htbp]
\centerline{\includegraphics[width=0.4\textwidth]{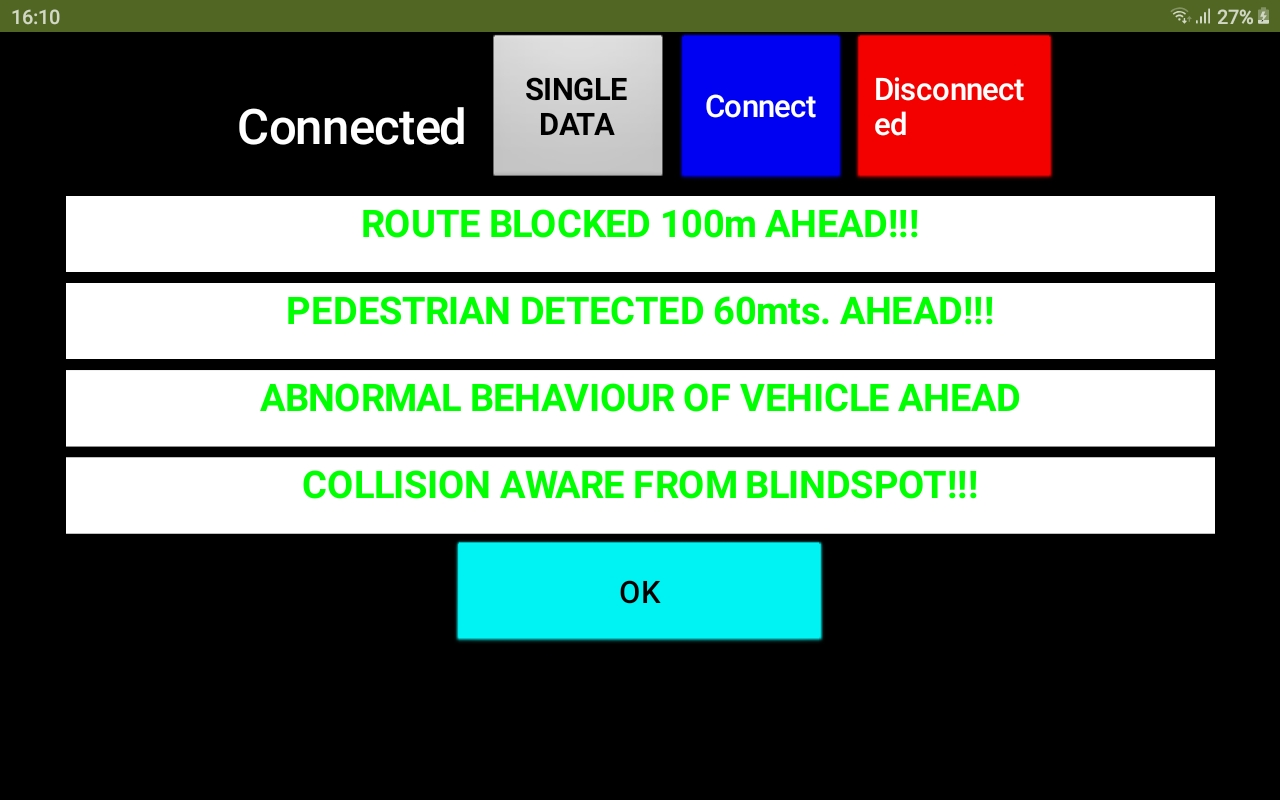}}
\caption{Android App for Driver assistance}
\label{fig}
\end{figure}

\subsection{Base Station Setup}
We have set up a centralized base station for a particular area which gives information about the surroundings of RSU's region. This base station gets data from RSU to the server via Ethernet connectivity from RSUs, as all RSUs have in built 4G modem and Ethernet connectivity. A Grafana-based dashboard has been developed which keeps track of vehicle's data passing by a certain RSU. This dashboard also gives information about the emergency vehicle approaching a certain area which can be further communicated to traffic managers for further action.
\begin{figure}[htbp]
\centerline{\includegraphics[width=0.5\textwidth]{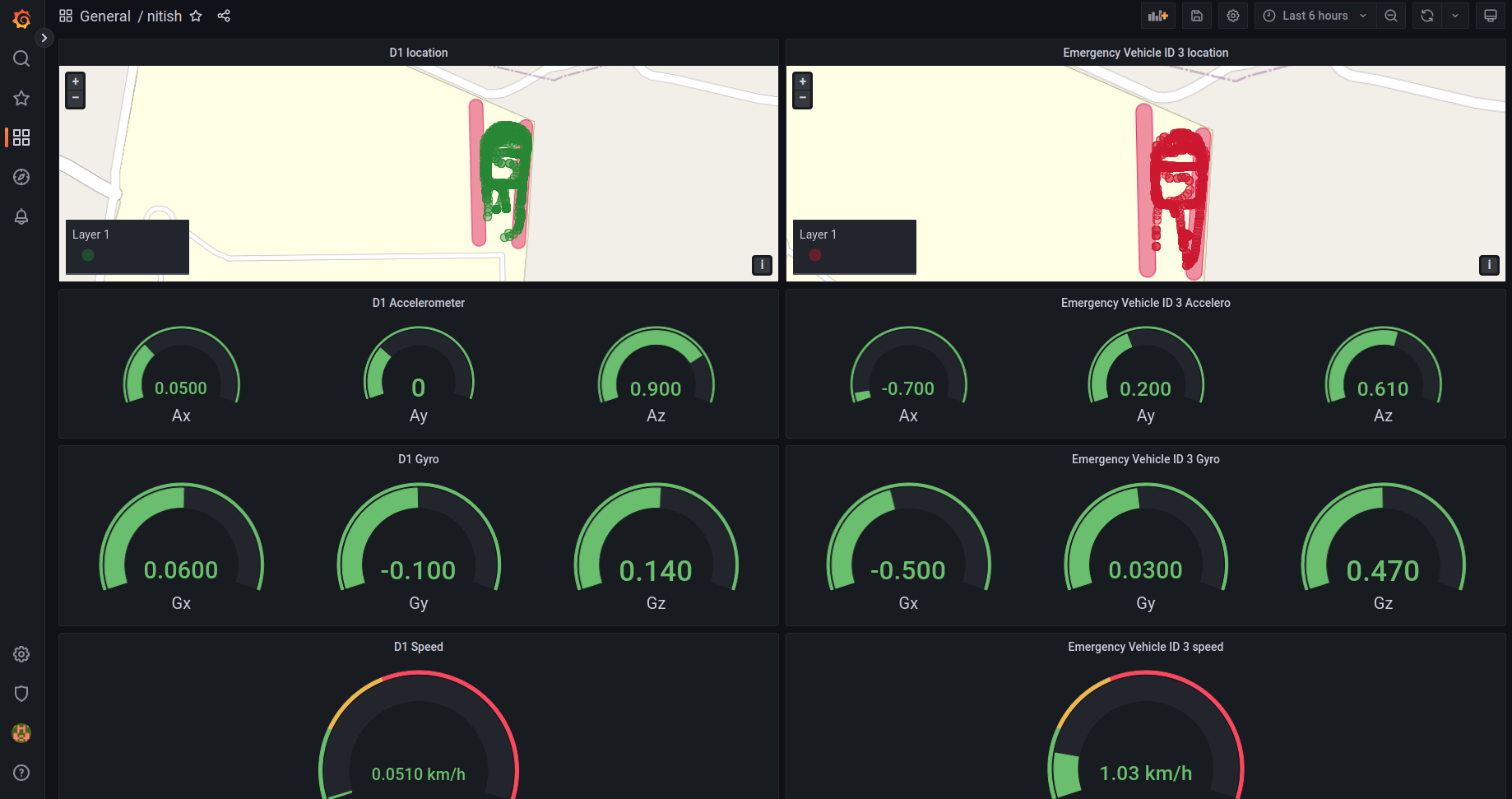}}
\caption{Grafana Dashboard for real-time monitoring of vehicles}
\label{fig}
\end{figure}

\section{Field Trials at TiHAN Testbed}
An interdisciplinary research project on smart mobility technologies called the Technology Innovation Hub on Autonomous Navigation (TiHAN) that aims to deploy applications for autonomous navigation was setup at The Indian Institute of Technology (IIT) Hyderabad. IIT Hyderabad campus features proving grounds, testing tracks, ground control stations, road infrastructure including smart poles, etc., was chosen as the location for our testing.
\begin{figure}[htbp]
\centerline{\includegraphics[width=0.5\textwidth]{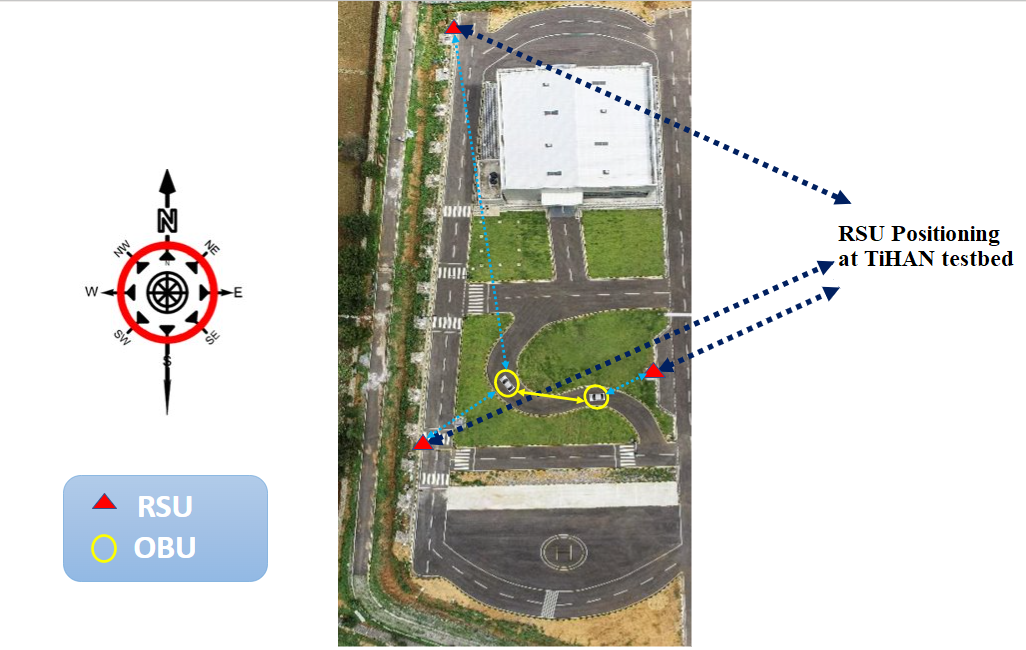}}
\caption{Testbed track at TiHAN with RSU and OBU Positioning}
\label{fig}
\end{figure}
\\
The DSRC applications are put to the test as part of our work by creating algorithms for various alert generations, which are as follows.
\subsection{Abnormal behavior of Vehicles around}
OBU has an inbuilt IMU in it which keeps on transmitting data such as accelerometer values, Gyroscope readings, etc of vehicles around in packets. After numerous trials and observations of these data, a threshold was set to accelerometer and gyroscope values from which we can detect any vehicle behaving abnormally. Abnormal behavior of vehicle means a vehicle is undergoing abrupt left and right turns with unusual speed. This behavior is recorded by OBU and is sent to App from which the driver of the emergency vehicle can take necessary actions.

\subsection{Accident or collision detection}
We have developed an algorithm which determines that whether there is a sudden break applied in the vehicle travelling ahead of the emergency vehicle or not. As there is a CAN interface in the OBU which gets all the vehicle's data to be processed in the OBU, here in the OBU we are also getting the speed of the vehicle at every time stamp. Target vehicle's speed is transmitted in every packet via DSRC antenna and this speed is stored in the form of an array in RSU which will be sent to emergency vehicle. Now consecutive differences of the speed of last seven packets are calculated everytime. If any vehicle applies an emergency break then there will a sudden rise in the consecutive difference. So, a threshold is set in this difference and when the difference is beyond a threshold value then an alert will be generated in the emergency vehicle and at nearby base station stating that a vehicle ahead has applied emergency break or accident has happened. This emergency break alert can be a sign of a collision or accident happening at that place. In certain situations usually, path blockage will be there which causes congestion and traffic jams. In such situations, traffic managers can broadcast a message from RSU to nearby OBUs to take an alternate path to avoid such congestion.

\subsection{Vulnerable road users (VRUs) Detection}
\begin{figure}[htbp]
\centerline{\includegraphics[width=0.5\textwidth]{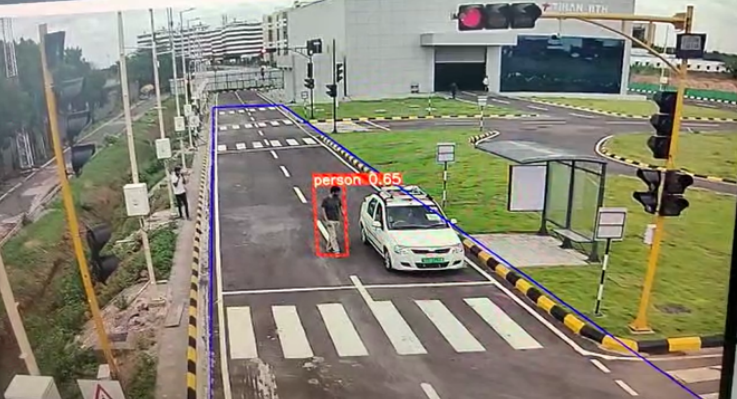}}
\caption{Pedestrian Detection using IP camera mounted near RSU}
\label{fig}
\end{figure}
We have an IP camera mounted on the smart poles which is used for security and monitoring purposes. These cameras are directly connected to the base station and can also be connected to RSUs. The most recent object identification model developed by ultralytics, the company that produced the Pytorch version of YOLOv3, is called YOLOv5, and it was released in June 2020. We have used YOLOv5 for pedestrian detection as pedestrian is also one of the class in YOLOv5 \cite{b19}. In YOLOv5, not only pedestrian but also we have classes for various vulnerable road users (VRUs) such as dogs, cows, cats, etc., that can also be detected and notified by the same camera. A code is written at the server end which connects our IP camera via real-time streaming protocol (RTSP) and this code runs continuously at the server end for pedestrian detection. The algorithm is developed such that, once a pedestrian is detected by an IP camera, and if the vehicle containing OBU is approaching the RSU region then emergency vehicle will immediately get an alert in the app stating that a pedestrian is on your pathway. This can avoid sudden accidents and save lives of many.

\subsection{Collision Avoidance or Blind Spot detection}
In some scenarios at certain intersection or at sharp turning it is very difficult to see a vehicle approaching from other intersection which is a blind spot at that turning point. We have implemented an algorithm in OBU which detects a vehicle at that blind spot and generates an alert stating that a vehicle is approaching at the same intersection.\\
\\
In this algorithm we predicted a possible collision point (x3,y3) at an intersection as shown in the figure 12.
\begin{figure}[htbp]
\centerline{\includegraphics[width=0.4\textwidth]{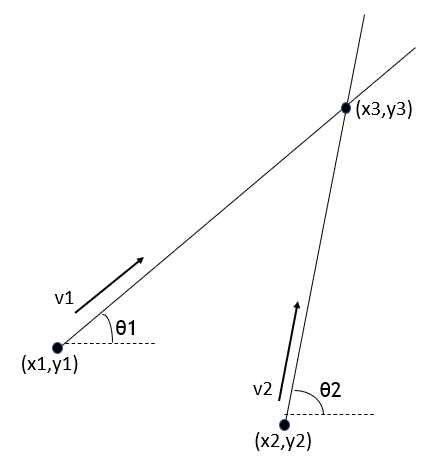}}
\caption{Intersection point of two vehicles}
\label{fig}
\end{figure}

Heading angles and coordinates of the vehicles [(x1,y1), $\theta1$ for emergency vehicle and (x2,y2), $\theta2$ for normal vehicle] are determined by GPS data. The algorithm involves the prediction of the possible collision point as per the formula given below \cite{b20}:
\[x3 = \frac{(y2-y1) - (x2*\tan \Theta 2 - x1*\tan \Theta 1)}{\tan \Theta 1 - \tan \Theta 2}\]\\
\[y3 = \frac{(x2-x1) - (y2*\cot \Theta 2 - y1*\cot \Theta 1)}{\cot \Theta 1 - \cot \Theta 2}\]
The distance between the possible collision point (predicted point) with the vehicles approaching that point is calculated with the appropriate heading angle. Now, if the two distances are of decreasing  nature and the difference between their heading angles is equal to the merging road angle then there might be a chance of collision at that intersection point. A threshold is setup based upon the decreasing distance, if the distance of two vehicles crosses that threshold then an immediate alert warning is sent to both drivers about collision alert so that they can be aware of the collision ahead.
\subsection{Alerts for Vehicles ahead of emergency vehicle}
As OBUs are mounted in vehicles moving on a straight road, one of them can be an emergency vehicle. At every 100 milliseconds, OBU broadcast the Basic Safety Messages (BSM), which provide their position, speed, heading, and acceleration. The OBUs receive the BSM messages whenever they enter the area and store them in a predetermined format. The algorithm operating in the OBU will calculate the vehicle's relative velocity if it is in front and indicate its position in relation to other vehicles (front, back, and sides). If the computed distance is less than the threshold value, which is currently set to 30m, OBU will send out an alert message to the vehicles ahead of the emergency vehicles. This warning message can be of the form that an emergency vehicle is approaching and is asking for a safe path from a commercial vehicle.
\section{Results and Discussions}
From the above alerts that has been developed and tested, can be combined to develop an application for free passage of emergency vehicle. Let us suppose an ambulance as our emergency vehicle which has OBU mounted on it. In case of huge traffic free passage of ambulance is a huge problem. Now we have developed a system which uses V2X technology to provide an easy path to the ambulance. The overall Architecture of our implementation is shown in figure 13.
\\
\begin{figure}[htbp]
\centerline{\includegraphics[width=0.5\textwidth]{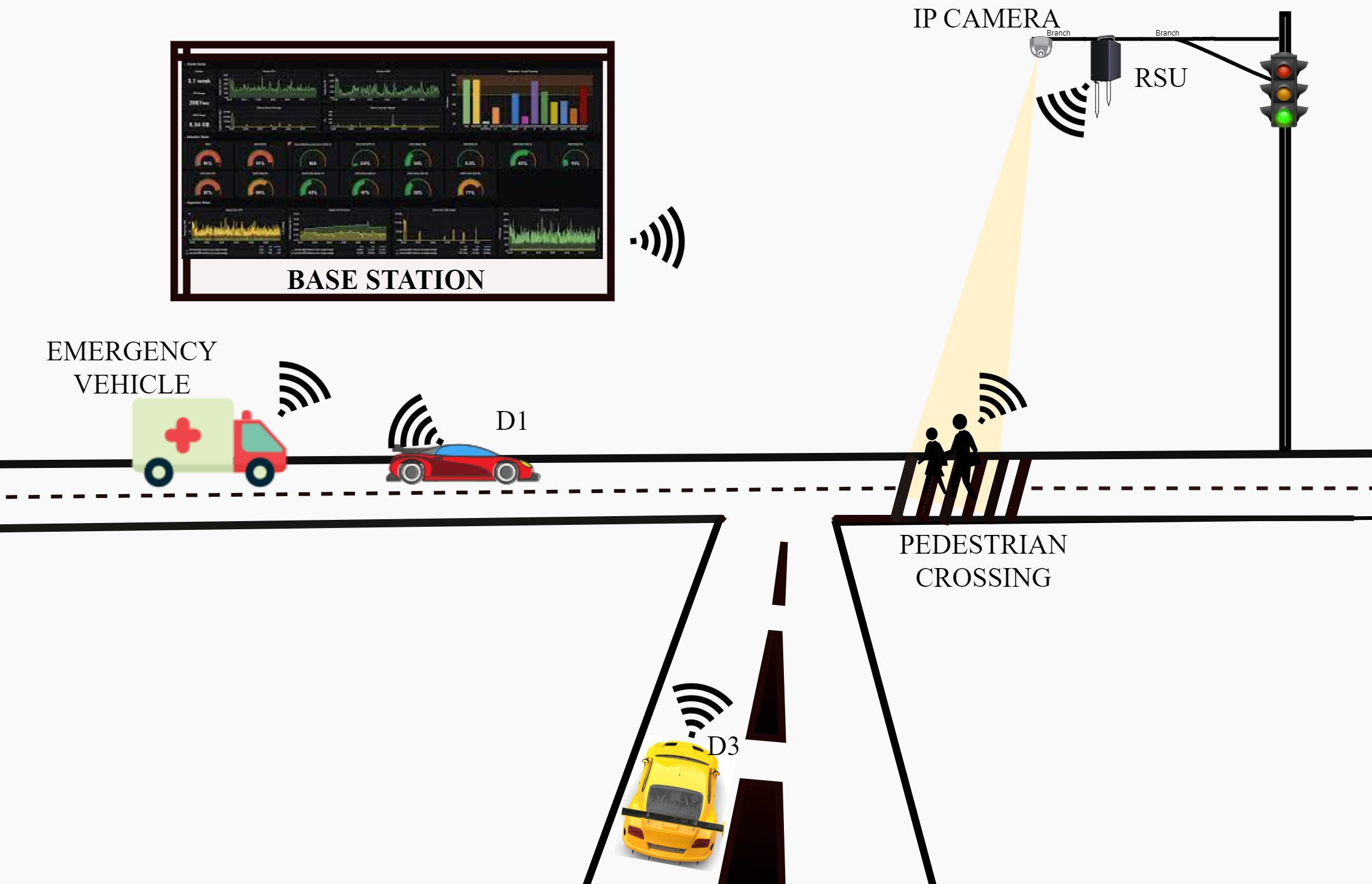}}
\caption{Overall Architecture of our Implementation}
\label{fig}
\end{figure}
\\
A program is written in the OBU such that it can receive all the sensor data information from other OBUs and RSUs, and transmits its own sensor data to nearby OBU and RSU. In the OBU all the above alert algorithms have been implemented and tested. So in case of huge traffic congestion, as ambulance arrives that area, RSU broadcast a safety message that an ambulance is arriving to all the OBUs nearby in that path. Hence, all commercial vehicles will get an alert beforehand that an emergency vehicle is approaching and to provide passage. The resultant alert is shown in figure 14.\\
\begin{figure}[htbp]
\centerline{\includegraphics[width=0.4\textwidth]{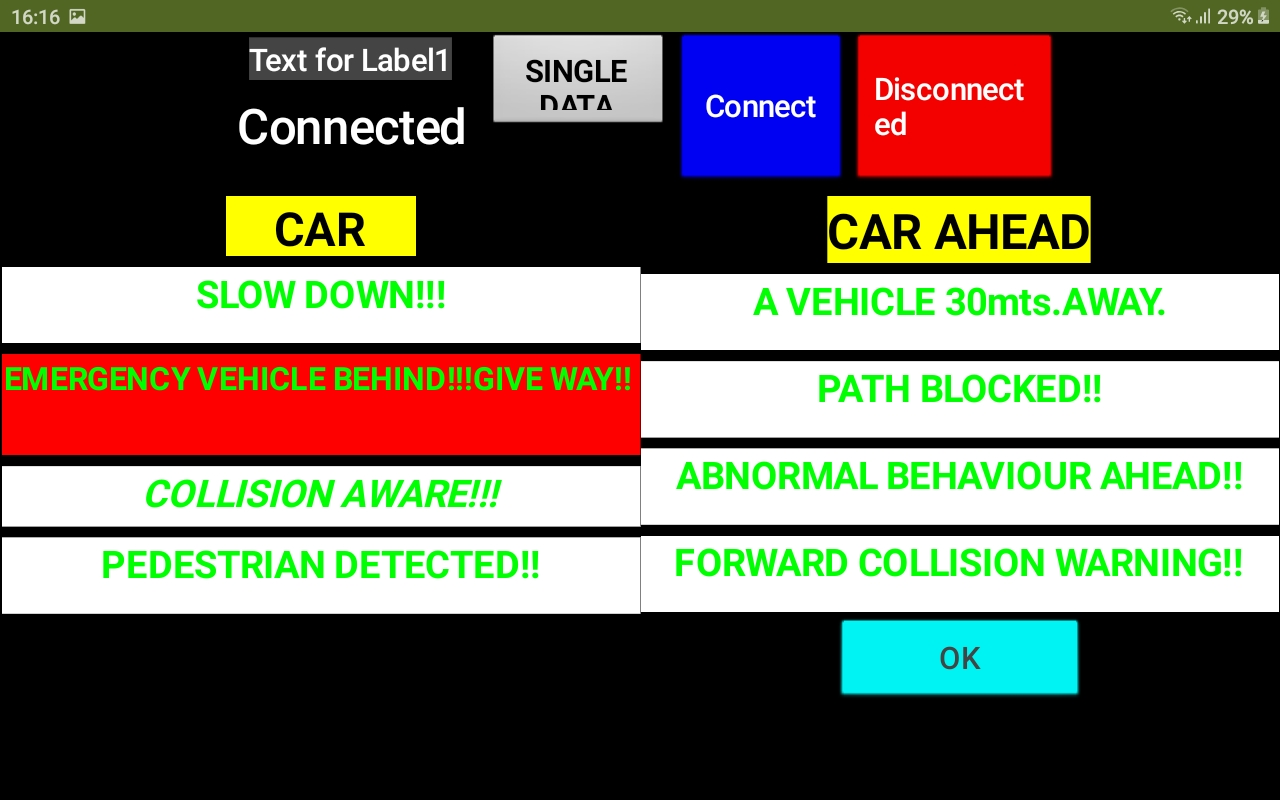}}
\caption{App for general vehicle getting an alert from emergency vehicle to provide pathway.}
\label{fig}
\end{figure}
\\
If any collision happened in the path of the Emergency vehicle usually that route is blocked. In that case due to accident or collision detection alert, RSU will broadcast this message to nearby OBUs about path blockage. This message when received by the OBU of emergency vehicle, the driver can take an alternate path to avoid delay due to path blockage. The resultant alert is shown in figure 15.\\
\begin{figure}[htbp]
\centerline{\includegraphics[width=0.4\textwidth]{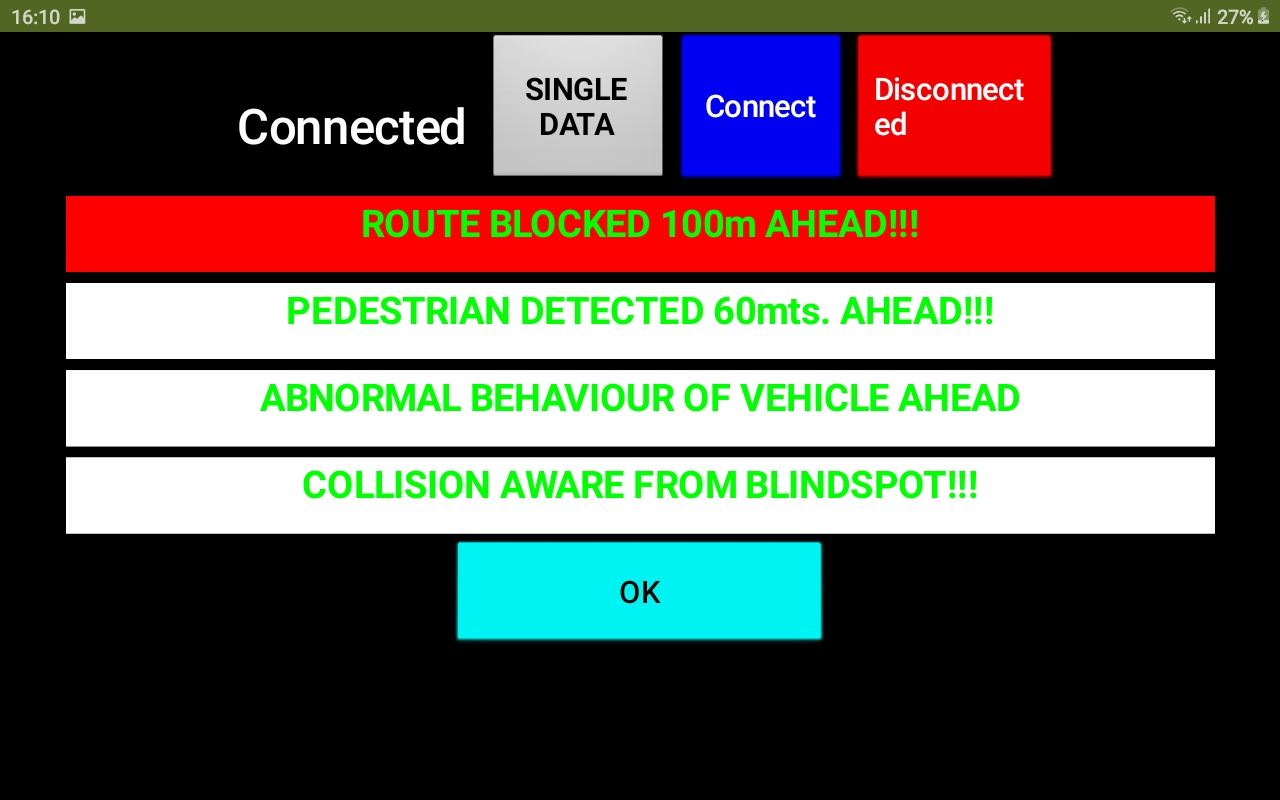}}
\caption{App for the Emergency vehicle getting an alert from RSU about blocked route ahead.}
\label{fig}
\end{figure}
\\
\\
Usually emergency vehicles travels at a faster speed which can be a threat to the pedestrian crossing nearby. So by pedestrian detection alert driver will get an alert that a pedestrian is crossing nearby in its path and hence driver can slow the vehicle at that moment. The resultant alert is shown in figure 16.
\begin{figure}[htbp]
\centerline{\includegraphics[width=0.4\textwidth]{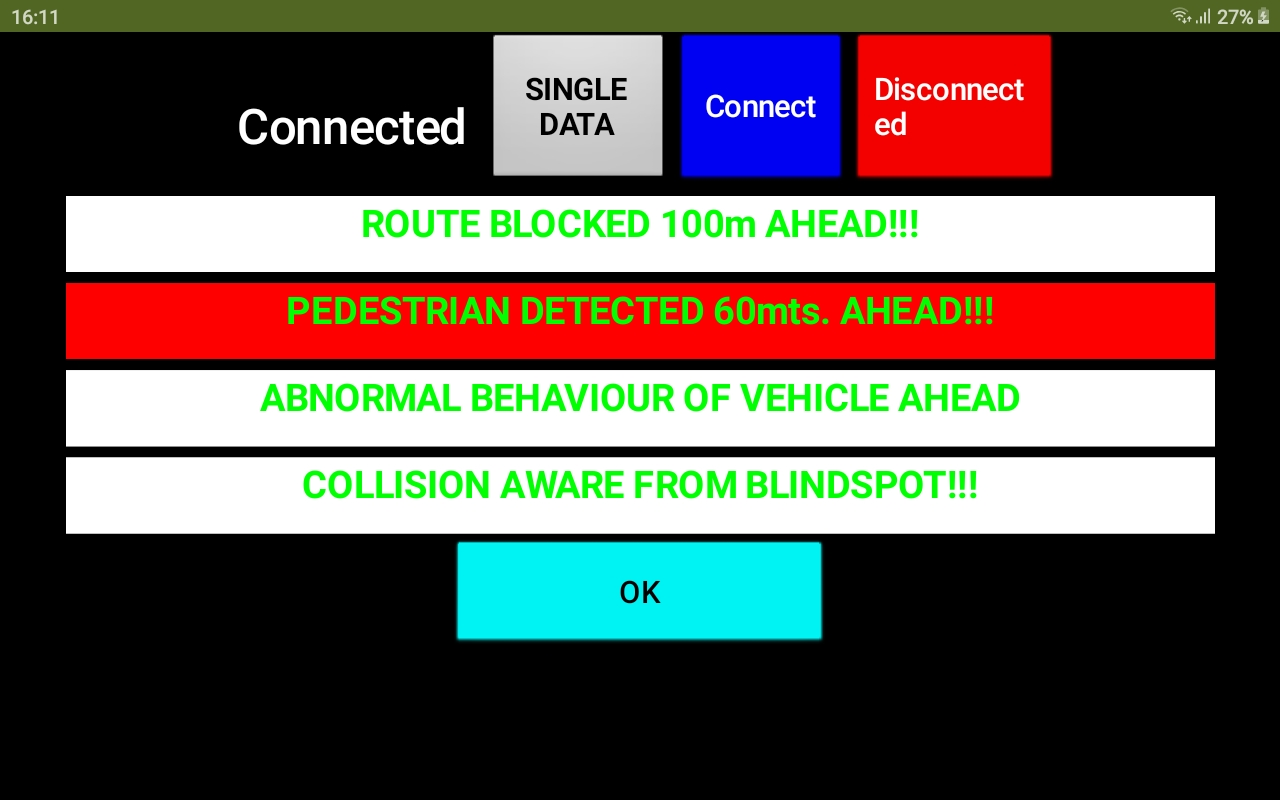}}
\caption{App for the Emergency vehicle getting an alert from RSU about pedestrian detection.}
\label{fig}
\end{figure}
Similarly, at blindspots, if any vehicle is approaching our emergency vehicle then from blindspot detection alert emergency vehicle will get to know about the approaching vehicle beforehand. The resultant alert is shown in figure 17.
\begin{figure}[htbp]
\centerline{\includegraphics[width=0.4\textwidth]{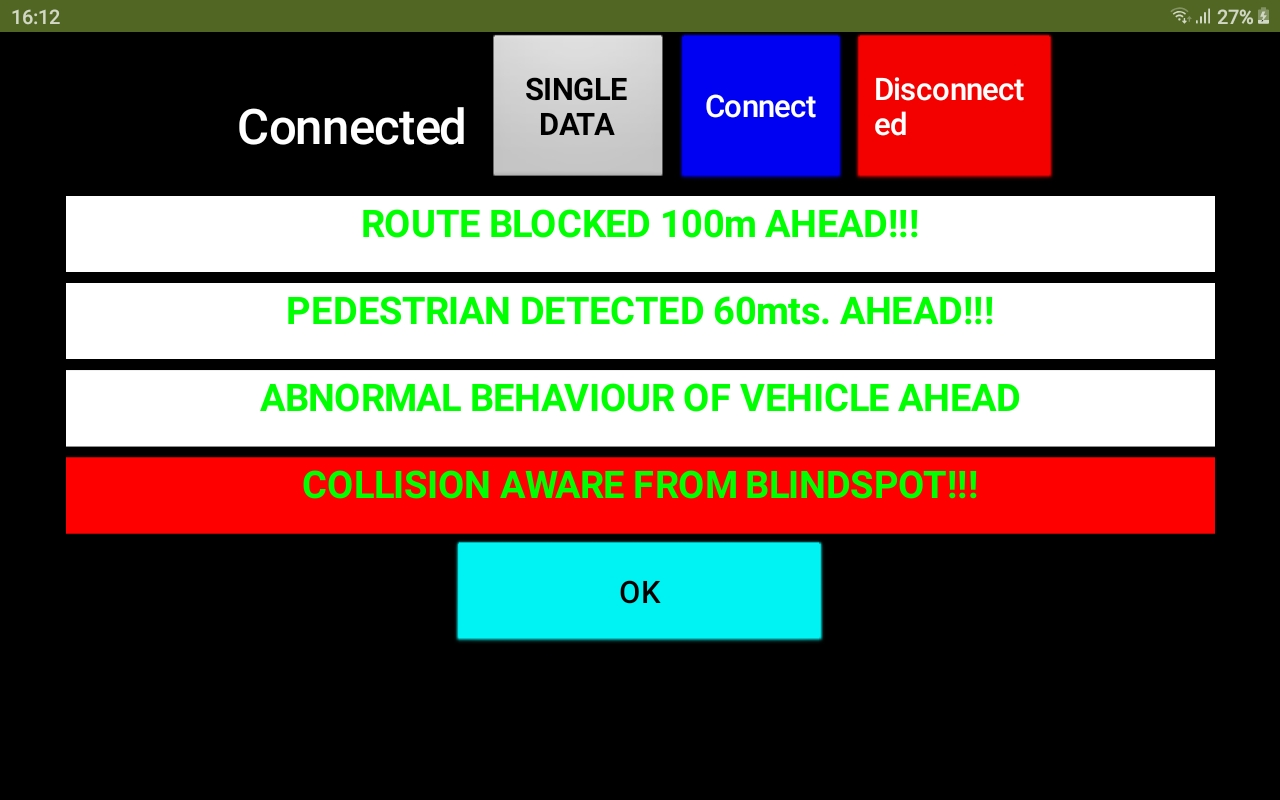}}
\caption{App for the Emergency vehicle getting an alert from OBU about collision awarness about the blindspot.}
\label{fig}
\end{figure}
\\
Also, there is a continuous real time monitoring system at the base station which keeps on providing information of the emergency vehicle to the base station. The grafana based dashboard can also run in smartphone's browser of the traffic manager so that they can take suitable action if any emergency vehicle is stuck somewhere due to some reason. 
\begin{figure}[htbp]
\centerline{\includegraphics[width=0.5\textwidth]{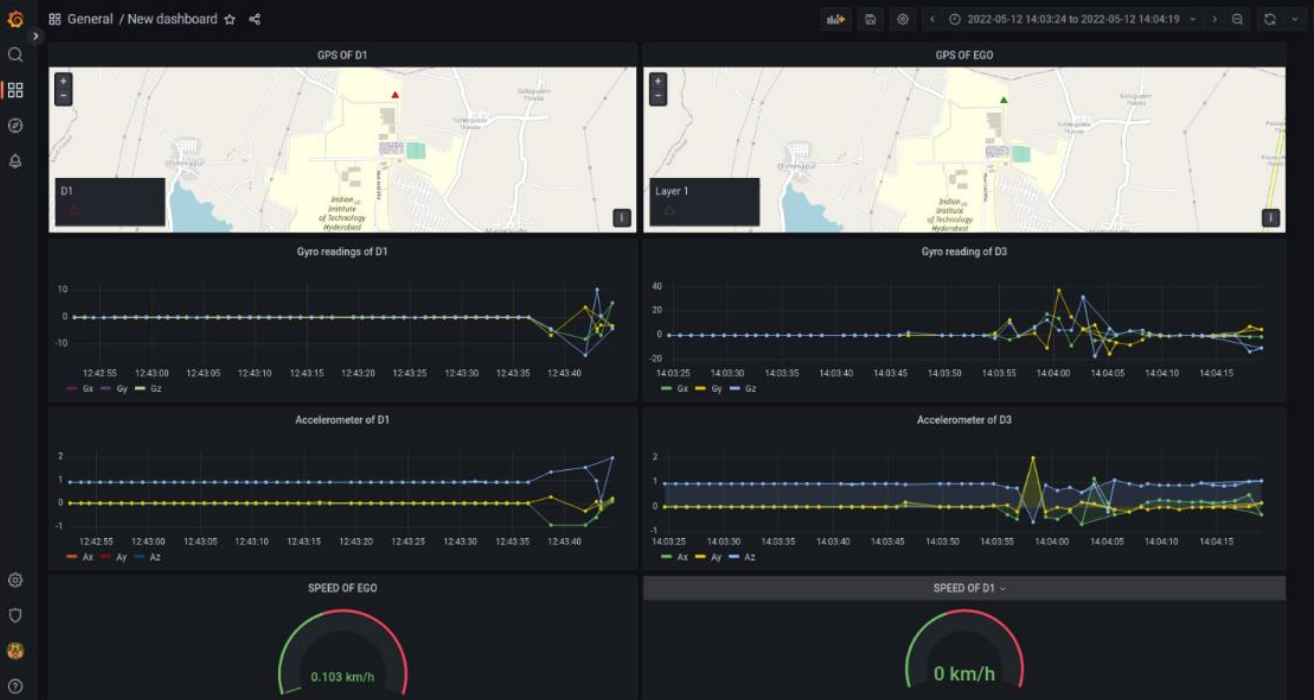}}
\caption{Live monitoring of vehicles using grafana dashboard.}
\label{fig}
\end{figure}
All the above use cases have been developed and tested at TiHAN testbed which gives promising results as shown above.
\section{Conclusion}
In this paper we have presented V2X communication with DSRC use cases which can be incorporated for smooth passage of Emergency vehicle. This paper focuses on the implementation of various algorithms in order to generate messages to alert the driver of the Emergency Vehicle about the surroundings with the help of data from the infrastructures and the other vehicles so that  the driver can act accordingly within time and prevent any miss-happening or delay in reaching the site which has been a major motivation of this research paper. We also came across a monitoring system that continuously monitors the vehicles in a particular region which can be used in case study of any accidents,miss-happenings and traffic rules violations. The overall architecture is developed for safety purpose and to give a new way for the vehicles to interact with each other.

\section{Acknowledgement}
This work was supported by DST National Mission Interdisciplinary Cyber-Physical Systems (NM-ICPS), Technology Innovation Hub on Autonomous Navigation and
Data Acquisition Systems: TiHAN Foundations at Indian
Institute of Technology (IIT) Hyderabad.

\end{document}